\documentstyle[twocolumn,epsfig]{article}
\newcommand{\beq}{\begin{equation}}
\newcommand{\eeq}{\end{equation}}

\def\fun#1#2{\lower3.6pt\vbox{\baselineskip0pt\lineskip.9pt
\ialign{$\mathsurround=0pt#1\hfil##\hfil$\crcr#2\crcr\sim\crcr}}}

\begin{document}
\twocolumn[

\begin{center}
{\Large \bf
RESONANT MESOMOLECULE FORMATION
}

\vspace*{0.3cm}
\large
Yu. V. Petrov and V. Yu. Petrov
\end{center}
]

Resonant formation of mesomolecules in the mixture of hydrogen
isotopes plays a key role in muon catalyzed fusion \cite{1}. In the
standard theory the rate of this process is calculated in the first
order of perturbation theory, and perturbation potential $\hat V$ is
calculated only in {\em dipole} approximation \cite{2,3}. The finite
width of the resonance \cite{4} and the anharmonism of the
potential \cite{5} contribute essentially to the result. In the
formation of the mesomolecule the most important perturbation is the
shift of the deuteron ion in respect to the centre of mass \cite{6,7}:
\beq
\hat V(\vec\rho,\vec{R}\ =\ U(|\vec{\rho}-\beta_2\vec{R}|)-U(\rho)\ .
\eeq
Here $U(\rho)$ is the $\mbox{}^1\Sigma_g$ term of the $DX$ mesomolecule
($X=H,D,T$); $\vec{\rho}$ is the distance between $d^+$ and $X^+$ ions,
$\vec{R}$ is the distance between mesoatom and $d^+$ inside the
mesomolecule, parameter $\beta_2$(which depends on mass ratios) is
$\beta_2$=0.5--$0.6$. The potential $U(\rho)$ is calculated in
Born-Oppenheimer approximation (infinite mass of nuclei) and
corrections to the finite nuclei mass are known to be small \cite{8}.

Dipole interaction is only the first term of the multipole expansion
for potential of Eq. (1). Meanwhile, the expansion parameter appears
to be not a small number. Indeed, this parameter is equal to the ratio
of $\beta_2R$ to vibrational amplitude $a=(M\Omega)^{-1/2}$ which
is $\beta_2R(M\Omega)^{1/2}\approx 0.2-0.35$. For this reason the next
(quadrupole) term of the multipole expansion changes formation rate
significantly.

In Ref. \cite{7} we developed a method which does not make use of the
multipole expansion --- the potential of Eq. (1) is taken into account
exactly. We do not need  the wave function of the mesomolecule,
it is enough to know only its asymptotics. This is due to the fact that
in the loosely bound mesomolecules $dy\mu$ ($v=J=1$; $y=d,t$) the
mesoatom $(y\mu)_{1s}$ is located in the average at large distances
from the deuteron and hence one can neglect their interaction. Only
exact binding energy $|\tilde{\varepsilon}_{11}|$ and constant
$C_{dy\mu}$ which enters the asymptotics of the wave function \cite{9}
are required for the calculation of formation rate. At the moment the
binding energy $|\tilde{\varepsilon}_{11}|$ is known with accuracy of
the order of 1~meV. The most accurate values of the constant
$C_{dy\mu}$ are obtained in Ref. \cite{10}.

On the basis of the method developed in \cite{7} we elaborated the
computer code designed for  the calculation of mesomolecule formation
rates \cite{11}. The subroutine MATEL of the code solves the
Schr\"odinger equation with potential $U(\rho)$ for $DX$-molecules and
mesomolecular complexes (MMCs). It determines the energy levels with
different vibrational ($\nu$) and rotational ($K$) quantum numbers and
also calculates their wave functions. On the next stage the code
calculates the matrix elements
$\langle\nu_i,K_i,F|\hat{V}|\nu_f,K_f,S\rangle$ where $F$ is the  spin
of the original mesoatom and $S$ is the spin of the mesomolecule. We
need to know something like 200 matrix elements and their dependence on
the energy. The code calculates also the position of all resonances
below and above the threshold. Next subroutine (SPEED) sums up all
Breit-Wigner resonance contributions to the probability of
mesomolecular complex formation $\lambda^{F\rightarrow S}_{\cal F}$ and
translates them from the centre-of-mass frame to the lab. one:
\beq
\lambda^{F\rightarrow S}_{\cal F}(E_L,T)=\int\! dE_c\,
\lambda^{F\rightarrow S}_{\cal F}(E_c)F(E_c\longrightarrow E_L,T).
\eeq
Here the function $F(E_c\longrightarrow E_L,T)$, which takes into
account the motion of gas molecules, leads to the Doppler broadening of
resonances with the width $\Delta_{DQ}=(4E_{QL}^rT/A)^{1/2}$ \cite{12}.
In the spirit of the Bohr theory of compound nuclei, we consider the
decay of mesomolecular complex as being independent of the channel by
which it was formed. Therefore in order to obtain the rate of formation
we have to multiply $\lambda^{F\rightarrow S}_{\cal F}(E_L,T)$ by  the
ratio $\Gamma_r/\Gamma_t$ where $\Gamma_t$ is the total and $\Gamma_r$
is the partial reaction width ($\Gamma_f$ for the fusion,
$\Gamma^{S\to F^\prime}_{BD}$ for the back decay, and so on).
One has to average the values of  $\lambda^{F}_{r}(E_L,T)$ over the
spectrum of mesoatoms $f(E_L)$.

For the mesoatom spin $F=0$ we calculated the rates of resonant
formation and fusion in reactions $(t\mu)^0+D_2$ and $(t\mu)^0+DH$ at
moderate gas density, when one can neglect rescattering of the MMC on
the gas molecules \cite{12}. In this case the decay of the MMC occurs
from the state where it was formed. At small $T$ the fusion rate is
determined by underthreshold resonance with the energy $E_r=-14$~meV:
\beq
\lambda^0_f(T)\ =\ B\frac{\Gamma_f}{E^2_\Gamma}\left|
\langle0,0,0|\hat{V}|2,1,1\rangle \right|^2\ ,
\eeq
that corresponds to the transition $\nu=0\rightarrow2$,
$K=0\to1$. Here $B=N_0\alpha(\hbar c)^2/m_e= 0.7085~\cdot~
10^{16}$eV/c, $\tilde{\Gamma}_f=0.65$~meV and
$\left|\langle i|V|f\rangle\right|^2=0.60\cdot 10^{-8}$ a.u. \cite{13}.
According to Eq. (3), the rate $\lambda_f^0$ does not depend on the
temperature $T$. Account for other resonances leads to small
variations of  $\lambda_f^0$ with temperature (see Fig. 1). This should
be compared with the strong temperature dependence for the rate of
$dd\mu$ formation (Fig. 2). Let us notice that, if one takes account of
resonances above the threshold only (dashed line in Fig. 1), then
the rate of formation restores its strong temperature dependence. Also
let us point out that dipole approximation overestimates the matrix
element of Eq. (1) more than 3 times \cite{7,12}.

We plot on the Fig. 1 the single known experimental point. For the first
time without any fitting parameter we reach the agreement with data on
the level of 15\% \cite{13}. The disagreement  of previous models with
data in the $dt\mu$ case was pointed out in a number of papers.

\begin{figure}
\centerline{\epsfxsize=7.cm\epsfbox{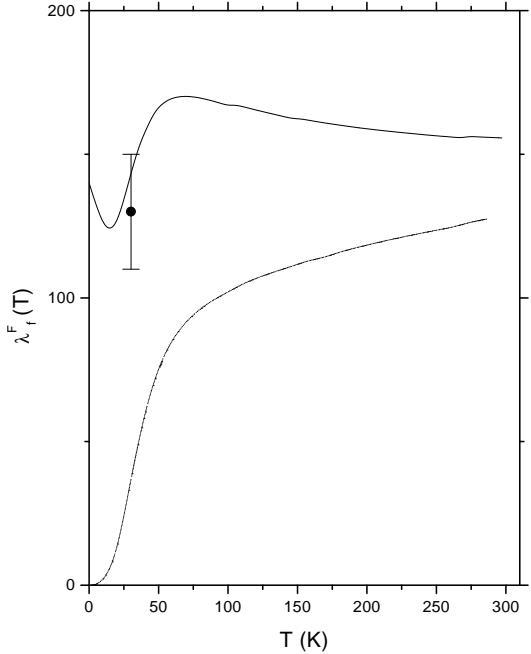}}
\caption{\footnotesize
 The rate of $dt\mu$ formation
with subsequent fusion $\lambda^0_f$ as a function of the temperature
$T$  at $F=0$ exact (solid curve) and in the approximation of zero
width of resonances \cite{13}. Experimental point was measured at
the density of equal to 1\% of the density of liquid hydrogen.
}
\end{figure}

\begin{figure}
\centerline{\epsfxsize=7.5cm\epsfbox{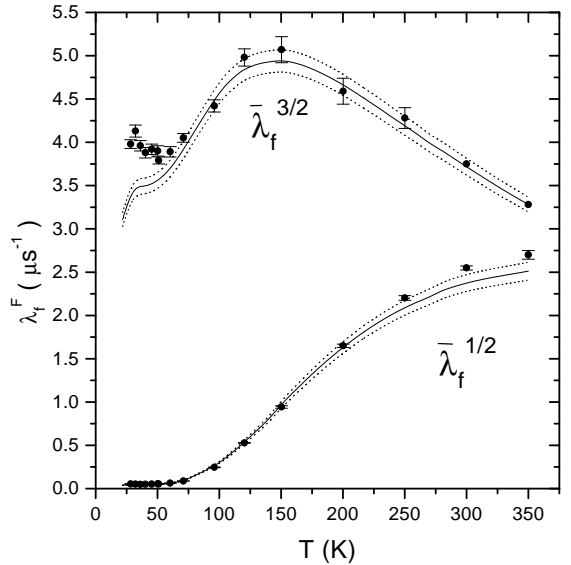}}
\caption{\footnotesize
Resonance rates of $dd\mu$ formation with subsequent
fusion for mesoatoms $(d\mu)_F$ with $F=1/2,3/2$. Solid curves
correspond to the theoretical calculations, dashed ones mark the
error corridor $\pm 4\%$  for $\lambda_f^{1/2}$ and
$\pm3\%$ for $\lambda_f^{3/2}$ \cite{15}.  Experimental points
were measured at the density  equal to $5\%$ of the density of
liquid hydrogen.
}
\end{figure}

The other limiting case when the collision width is much larger than all
other widths takes place for the resonant formation of
$dd\mu$-mesomolecules in $D_2$-gas \cite{14}. After formation of the
mesomolecular complex in the state with rotational quantum number
$K_f$ it can be changed due to collisions to some new $K_f^\prime$.
It is enough time to reach the equilibrium Boltzman distribution in
rotational quantum numbers $K_f^\prime$.
For this reason the back decay of the mesomolecular complex takes
place independently of the value $K_f$ with which it was formed.
Rates of the $dd\mu$ formation with subsequent fusion for two values of
the mesomolecule spin $F=1/2,3/2$ are presented on Fig.2 \cite{15}.
They are considerably less than for $dt\mu$ case. This fact is a
consequence of the sharp decrease of transition matrix elements with
the energy of the bound state ($\bar{\varepsilon}=-0.632\,$eV for
$dt\mu$ and $\bar{\varepsilon}=-1.966\,$eV for $dd\mu$).
The rate
$\lambda^{3/2}_f(T)$ depends weakly on $T$ owing to the strong Doppler
broadening of the $K=0\rightarrow 1$ resonance with small positive
resonance energy $E^r_c=4.2$ meV. On the contrary, the rate
$\lambda^{1/2}_f(T)$ (which is mainly due to the far
resonances with positive energy ) depends on the temperature rather strongly.

At Fig. 2 we plot also recent experimental data of Ref. \cite{16} which
were obtained for the non-equilibrium mixture of ortho- and
para-deuterium in the ratio of 2 to 1 (independent of $T$).
In such a
non-equilibrium gas due to the extra inelastic acceleration of
mesoatoms on para-deuteuriums (which are excited by $7.4\,$meV), the
mesoatom spectrum $f(E_L)$ below $\approx70\,$K should decrease
with $E_L$ considerably faster than the equilibrium Maxwell spectrum
$f_M(E_L)$. Meanwhile in our paper \cite{15} the fusion rate
of Eq. (2) was averaged over $f_M(E_L)$ spectrum.
For this reason it makes sense to compare the results of Ref. \cite{15}
with data only at temperatures $T>70\,$K.
The calculation (taking into account non-resonant formation of $dd\mu$
\cite{17}) based on purely theoretical values of all quantities,
reproduces the data on the level of 10\%. The accuracy of our
calculations is limited mainly by insufficient accuracy of the
estimation of fusion width $\Gamma_f=460(70)\,\mu\sec^{-1}$ \cite{18}.
If one uses $\Gamma_f$ as a single fitting parameter it is possible to
achieve an excellent relative accuracy of 3\% for $\lambda^{3/2}_f(T)$
and 4\% for $\lambda^{1/2}_f(T)$.
The best fit corresponds to $\Gamma_f=397(40)\,\mu\sec^{-1}$. This
value lies within the errors of theoretical estimate.

Let us compare our results with calculations of Refs. \cite{16,19}.
These calculations coincide with the data only if one introduces an
additional fitting parameter $|C_{matr}|^2=0.56$ which is used as factor
to multiply all transition matrix elements squared. In other words,
this approach overestimates $\lambda^F_f$ by a factor of $1.8$
(see Table 1). In our calculations this fitting parameter is absent.
The discrepancy between the data and our calculations in the region
of $T<70\,K$ (up to $14$\% at $T=28.3\,$K) can be, in our opinion,
attributed to the distortion of the spectrum $f(E_L)$
due to the mesoatom acceleration in non-equilibrium medium.
This new effect awaits for calculation.

Thus the theory of the mesomolecule resonant formation
\cite{2}--\cite{4} by Vesman mechanism developed in Refs.
\cite{2}--\cite{4} and improved significantly in Refs \cite{7,11,12}
describes now the data with sufficient accuracy (see Figs. 1,2).

\begin{table}
\caption{}
\begin{center}
\begin{tabular}{|c|c|c|}
\hline
& $|C_{matr}|^2$ & $\tilde{\Gamma}_f$, $\mu$sec$^{-1}$\\
\hline
IAE, 1993$^{1)}$ \cite{19}& 0.62(13) & 337(40) \\
\hline
IAE, 2001 \cite{16} & 0.56(3) & 407(20)   \\
\hline
PNPI, 1998 \cite{15}& 1 & 400(46)  \\
\hline
PNPI, 2001 & 1 & 397(40)  \\
\hline
\end{tabular}
\end{center}
\end{table}


\begin{thebibliography}{99}
\bibitem{1}
S.S. Gerstein, Yu.V.Petrov and L.I.~Ponomarev,  Usp.Fiz.Nauk
{\bf160} (1990) 3;  Sov.Phys.Usp.\\ {\bf33} (1990) 591.

\bibitem{2} S.S. Gerstein and L.I.~Ponomarev, Phys. Lett. {\bf72}B
(1977) 80.

\bibitem{3} M.P. Faifman, L.I.~Menshikov and T.A.~Strizh,
Muon Cat. Fus. {\bf4} (1989) 1.

\bibitem{4} Yu.V. Petrov,  Phys. Lett. B{\bf163}(1985) 28.

\bibitem{5} Yu.V.~Petrov, V.Yu.~Petrov and
A.I.~Schlyakhter, Muon Cat. Fus. {\bf2} (1988) 261.

\bibitem{6} A. Scrinzi,  Muon Cat. Fus. {\bf5/6} (1990/91) 179.

\bibitem{7}
Yu.V.~Petrov and V.Yu.~Petrov,  ZhETF
{\bf100} (1991) 56;  JETP {\bf73} (1991) 29.

\bibitem{8} W.~Kolos and L.J.~Wolniewicz, Chem.Phys. {\bf43}(1965) 2429.

\bibitem{9}
L. Men'shikov,  Yad. Fiz. {\bf42} (1985) 1184;\\
 Sov. J. Nucl. Phys. {\bf42} (1985) 750.

\bibitem{10}
Y. Kino {\em et al}, Hyperfine Interactions\\ {\bf 101/102} (1996) 325.

\bibitem{11}
V.Yu. Petrov, {\em The code MATEL and SPEED} PNPI, Gatchina, 1992.

\bibitem{12}
Yu.V.~Petrov, V.Yu.~Petrov, H.H.~Schmidt, Phys. Lett.  {\bf
331} (1994) 266.

\bibitem{13}
Yu.V. Petrov and V.Yu. Petrov,  Phys. Lett.\\
B{\bf378} (1996) 1.

\bibitem{14}
V.V.~Kuzminov, Yu.~V.Petrov and V.Yu.~Petrov,
 Hyperfine Interactions {\bf101/102} (1996) 197.

\bibitem{15}
V.V.~Kuzminov, Yu.V.~Petrov, V.Yu.~Pe\-trov,
W.H.~Bre\"unlich {\em et al.\/},
 Phys. Rev. A{\bf57} (1998) 1636.

\bibitem{16}
N.I.~Voropaev,~D.V.~Balin, W.H.~Bre\"un\-lich {\em et al.\/},
{\em"Final results of $\mu CF$ experiment in $D_2$ and $HD$}, in print.

\bibitem{17}
M.P. Faifman {\em et al.}, Muon Catalyzed Fusion\\ {\bf4} (1989) 341.

\bibitem{18}
L.N.~Bogdanova  {\it et al.}, Phys.Lett. {\bf 115}B (1982) 171;
{\bf167}B (1986) 485.

\bibitem{19}
A.~Scrinzi, P.~Kammel, J.~Zmes\-kal, W.H.~Bre\"un\-lich {\em et al},
Phys. Rev. A{\bf47} (1993) 4691.
\end{thebibliography}
\end{document}